\documentclass[manuscript]{acmart}
\AtBeginDocument{%
  }

\setcopyright{acmlicensed}
\copyrightyear{2018}
\acmYear{2018}
\acmDOI{XXXXXXX.XXXXXXX}
\acmConference[Conference acronym 'XX]{Make sure to enter the correct
  conference title from your rights confirmation email}{June 03--05,
  2018}{Woodstock, NY}

\acmISBN{978-1-4503-XXXX-X/2018/06}

\usepackage{booktabs}
\usepackage{multirow}
\usepackage{caption}
\usepackage{subcaption}
\usepackage{graphicx}
\usepackage{float}
\usepackage{rotating}
\usepackage{ragged2e}
\usepackage{tabularx}
\usepackage{courier}
\usepackage{pdfpages}
\usepackage{enumitem}
\usepackage{longtable}
\usepackage{array}
\usepackage{listings}
\usepackage{xcolor}
\definecolor{jsonpunct}{rgb}{0.40,0.40,0.40}
\usepackage{makecell}
\usepackage{float}
\usepackage{placeins}
\usepackage{adjustbox}
\usepackage{xurl}

\usepackage{pgfplots}
\pgfplotsset{compat=1.18}
\usepackage{pgfplotstable}
\usepgfplotslibrary{dateplot}
\usepgfplotslibrary{groupplots}
\usetikzlibrary{pgfplots.statistics}
\usetikzlibrary{pgfplots.groupplots}
\usetikzlibrary{matrix}
\usepackage{bbding}
\usepackage{svg}

\pgfplotsset{
    IDEA box/.style={
        width=\columnwidth,
        height=6cm,
        font=\sffamily,
        axis line style=very thin,
        separate axis lines,
        axis lines*=left,
        scaled y ticks=false,
        major tick length=0,
        major x tick style=transparent,
        enlarge x limits={abs=0.5cm},
        xlabel style={name=xlabel},
        ymajorgrids=true,
        ybar=0pt,
        clip=false,
        boxplot/draw direction=y,
        boxplot/whisker extend=0,
        boxplot/every box/.style={fill=gray!50,draw=black},
        boxplot/every outlier/.style={mark=o,mark size=1.5pt,black,fill=white}
        boxplot/every median/.style={thick,fill=black,draw=black},
        boxplot/draw/whisker/.code 2 args={\draw[/pgfplots/boxplot/every whisker/.try] (boxplot cs:##1) -- (boxplot cs:##2);},
        boxplot/every whisker/.style={very thick},
        x tick label style={font=\footnotesize, align=center},
        cycle list={},
    }
}

\begin{document}
\title{DeTAILS: Deep Thematic Analysis with Iterative LLM Support}

\author{Ansh Sharma}
\email{a646shar@uwaterloo.ca}
\orcid{0000-0003-1892-0668}
\affiliation{%
  \institution{University of Waterloo}
  \city{Waterloo}
  \state{Ontario}
  \country{Canada}
}

\author{Karen Cochrane}
\email{karen.cochrane@uwaterloo.ca}
\orcid{0000-0002-4563-0618}
\affiliation{%
  \institution{University of Waterloo}
  \city{Waterloo}
  \state{Ontario}
  \country{Canada}
}

\author{James R. Wallace}
\email{james.wallace@uwaterloo.ca}
\orcid{0000-0002-5162-0256}
\affiliation{%
  \institution{University of Waterloo}
  \city{Waterloo}
  \state{Ontario}
  \country{Canada}
}

\renewcommand{\shortauthors}{Sharma et al.}

\begin{abstract}
Thematic analysis is widely used in qualitative research but can be difficult to scale because of its iterative, interpretive demands. We introduce DeTAILS, a toolkit that integrates large language model (LLM) assistance into a workflow inspired by Braun and Clarke’s thematic analysis framework. DeTAILS supports researchers in generating and refining codes, reviewing clusters, and synthesizing themes through interactive feedback loops designed to preserve analytic agency. We evaluated the system with 18 qualitative researchers analyzing Reddit data. Quantitative results showed strong alignment between LLM-supported outputs and participants’ refinements, alongside reduced workload and high perceived usefulness. Qualitatively, participants reported that DeTAILS accelerated analysis, prompted reflexive engagement with AI outputs, and fostered trust through transparency and control. We contribute: (1) an interactive human–LLM workflow for large-scale qualitative analysis, (2) empirical evidence of its feasibility and researcher experience, and (3) design implications for trustworthy AI-assisted qualitative research.
\end{abstract}

\maketitle

\section{Introduction}

The growing scale of online datasets, such as social media posts, has created a pressing need for methods that can support efficient yet rigorous analysis~\cite{abhinaya2024bigqual}. Qualitative research methods such as thematic analysis (TA)~\cite{BraunANDClarke2006} enable scholars to understand meanings, experiences, and perspectives, producing rich insights into complex social phenomena~\cite{denzin2011sage} and situating findings within broader cultural or situational contexts~\cite{creswell2016qualitative}. These approaches are particularly valuable for describing behaviour, exploring lived experience, and identifying concepts or variables that might otherwise be overlooked~\cite{Rice1996, patton2014qualitative}. However, despite these strengths, TA is time- and labour-intensive, and scaling it to large corpora presents practical challenges for consistency and depth~\cite{BraunANDClarke2021, Guest2012, jiang2021serendipity}.

Importantly, TA is not a single technique but a family of approaches~\cite{creswell2016qualitative}. Reflexive Thematic Analysis (RTA), developed by Braun and Clarke, emphasizes the researcher’s interpretive role, the evolving and flexible nature of codes, and reflexivity as a way to acknowledge how analytic decisions shape findings~\cite{BraunANDClarke2019, ayton2023thematic}. In contrast, codebook TA stabilizes codes into more structured frameworks, supporting consistency, rigour, and team-based analysis~\cite{Boyatzis1998,roberts2019attempting}. Both approaches are widely used, but they carry different commitments to reflexivity, flexibility, and replicability, which become particularly salient when scaling analysis to large datasets.

Large language models (LLMs) have recently been proposed as tools to help meet these scaling challenges. Their ability to generate, classify, and summarize text makes them promising assistants for coding and theme development~\cite{deiner2024inductivetheme, tai2024aidanalysis}. Yet existing AI-assisted qualitative tools often produce outputs that are generic or decontextualized, require extensive refinement, and operate through opaque processes that undermine trust~\cite{Burrell2016, Feuston2021, jiang2021serendipity}. Analysts frequently prefer manual methods when reflexivity and interpretive authority are central~\cite{BraunANDClarke2019}. As a result, despite the promise of LLMs, their adoption in qualitative research remains limited, with open questions around how to balance efficiency with transparency, trust, and interpretive depth.

To address these challenges, we developed \emph{DeTAILS: Deep Thematic Analysis with Iterative LLM Support}, a researcher-centred toolkit inspired by Braun and Clarke’s six phases of analysis. Although originally motivated by RTA, the scale of Reddit datasets required design choices such as editable but stabilized codebooks, which ultimately aligned the workflow more closely with codebook TA~\cite{Boyatzis1998,roberts2019attempting}. Reflecting on our evaluation, however, we observed that reflexivity remained central: participants frequently verified, critiqued, and reworked AI outputs, treating this engagement as an integral part of their analysis.

Our evaluation with 18 qualitative researchers demonstrated both the potential and the tensions of this approach. DeTAILS substantially accelerated coding and theme development and was valued for its structured workflow and support for researcher control. At the same time, participants noted areas for improvement: reflexivity often centred on verifying AI outputs rather than interrogating analytic assumptions, novices at times over-trusted fluent outputs, and the workflow risked constraining interpretive flexibility. Through our own analysis of these interviews, using a codebook-style approach similar to what DeTAILS ultimately supported, we developed themes that informed a set of design recommendations for future iterations. These findings showcase the system’s promise while also highlighting what needs to change in order to better align with reflexive thematic analysis.

This paper makes three contributions to HCI and qualitative research. First, it introduces DeTAILS, an interactive toolkit that integrates LLM support across thematic analysis phases while balancing acceleration with researcher control. Second, it provides a nuanced empirical account of how qualitative researchers of varying expertise engaged with LLM assistance, highlighting both efficiencies and tensions around agency, reflexivity, and trust. Third, it offers design implications for the next generation of AI-assisted qualitative tools, including opt-in AI involvement, iterative workflows that support both forward and backward reflection, and safeguards against over-trust. In doing so, this work addresses current gaps in how AI systems are applied to qualitative research, where existing tools often privilege efficiency over reflexivity, provide opaque or decontextualised outputs, and leave unanswered questions about trust and adoption. Our findings demonstrate both the potential and the limits of current approaches, and chart a path toward systems that scale qualitative analysis without eroding its interpretive commitments.

\section{Related Work}
Thematic Analysis (TA) is a foundational qualitative method for identifying patterns of meaning (“themes”) within textual data~\cite{Jennifer-Attride-Stirling2001, Boyatzis1998, BraunANDClarke2006}. Reflexive TA (RTA), developed by Braun and Clarke~\cite{BraunANDClarke2019}, has become especially influential. In RTA, the researcher plays an active, interpretive role in theme development rather than applying a fixed codebook or pursuing inter-coder reliability~\cite{terry2017}. Codes and themes are not pre-defined but emerge through close engagement with the data and ongoing critical reflection~\cite{BraunANDClarke2006, Nowell2017, saldana2021}. Reflexivity is central: researchers acknowledge their subjectivity and positionality, recognising how analytic decisions shape knowledge production~\cite{LINCOLN1985, Morrow2005}. This interpretive freedom allows for the generation of rich, situated insights but also demands rigour in documenting decisions and engaging critically with one’s own role in the analysis~\cite{Finlay2002, LINCOLN1985, Nowell2017}.

These qualities made RTA an attractive starting point for our work. We sought to design a system that could preserve interpretive depth and reflexivity while helping researchers handle larger, more complex corpora such as social media data. Yet RTA is inherently time- and labour-intensive, particularly at scale~\cite{Guest2012, Nelson2017, Nowell2017}. Researchers must develop deep familiarity with their material and iteratively revisit analytic decisions, which becomes impractical with thousands of posts or transcripts~\cite{Guest2012, Nowell2017, saldana2021}. Collaborative projects further add overhead as teams negotiate divergent perspectives and reconcile competing theme definitions~\cite{CASTLEBERRY2018, Guest2012, MacPhail2016}. Without large teams or extended timelines, analysts risk producing superficial themes or overlooking patterns in the data~\cite{BraunANDClarke2021, Guest2012, Tracy2010}. For example, platforms such as Reddit can generate corpora so vast that manual reflexive engagement is nearly impossible~\cite{baumgartner2020, gauthier2022}.

Although computational tools exist for tasks such as text clustering or keyword extraction~\cite{Blei2003, Jockers2013, Nelson2020}, few align with the interpretive and reflexive commitments of RTA~\cite{Baumer2017, Muller2016, Kaur2024}. Prior work on computational grounded theory and mixed-methods text analysis demonstrates how algorithmic techniques can surface large-scale patterns, but cautions that such tools often lack transparency and can encourage superficial engagement with data~\cite{Nelson2020, Baumer2017, Muller2016}. Very few systems combine algorithmic support with features that foreground researcher oversight, iterative refinement, and reflexivity. Our work was motivated by this gap: we aimed to explore how LLMs could be embedded within an interactive workflow that supports the interpretive and reflexive principles of RTA while making large-scale qualitative analysis more tractable.

\subsection{Large Language Models as Qualitative Coding Assistants}

Large language models (LLMs) are transformer-based neural networks trained on extensive corpora to generate, classify, and summarize text~\cite{eloundou2024llmsurvey,openai2023gpt4,zhao2023surveyllm}. Because they produce fluent, contextually relevant prose and can respond conversationally, they appear well suited to tasks such as suggesting candidate codes or drafting theme summaries. These capabilities have prompted interest in computational tools that assist human analysts~\cite{graber2017,gauthier2022,Gillies2022}, and HCI researchers have begun to explore LLMs as qualitative coding assistants~\cite{dai2023llminloop,deiner2024inductivetheme,drapal2023supportthematicanalysis,gao2024mindcoder,zeyu2024dataannotation,Hou2024DeductiveCoding,Sun2024PsychotherapyCodes,Zhang2025ChatGPTRedesign}.

Despite promising results, studies report notable limitations in both deductive and inductive coding. In deductive workflows, few-shot prompting helps align outputs with a predefined codebook but often yields overgeneralized or incorrect labels~\cite{Chew2023Deductive,Gamieldien2023,Kirsten2024Complexity,klyukvin2024enhancingqualresearch,Meng2024Synergy,Rao2025QuaLLM,Sinha2024GroundedTheory,xiao2023combiningcodebook}. For inductive coding, LLMs can quickly produce plausible code lists and theme summaries, yet their suggestions depend heavily on prompt design~\cite{Zhang2023QualiGPT,depaoli2024inductive} and still require analysts to verify, refine, and interpret the outputs~\cite{Katz2024Codebook}. Widely documented issues---including hallucinations, overconfidence, shortcut reasoning, and generic explanations---also risk undermining analytic rigour~\cite{chen2024dataljustice,savelka2023biasqualanalysis,sun2023surveyhallucination,xu2024hallucinationinevitable,gao2024collabcoder,simret2023patat,koyejo2023trustworthyllms,sun2024trustllm,baker2025monitoringreasoningmodels,bommasani2022opportunitiesrisksfoundationmodels,Long2024AIWorkflowUtility,chen2023sensitivities,takagi2023prompttemplate,nyaaba2025optimizinggenerativeaisaccuracy}. These challenges underscore the need for careful human oversight~\cite{baker2025monitoringreasoningmodels}.

To address these concerns, recent HCI work has proposed human-in-the-loop workflows that foreground researcher judgment~\cite{xu2023aihci,GauthierIWNDWYT,gauthier2022}. \citet{Long2024AIWorkflowUtility} advocate AI workflow design patterns that interleave automated suggestions with deliberate checkpoints where users vet and approve outputs. CollabCoder exemplifies this approach by embedding LLM suggestions within a collaborative coding interface: analysts independently code the data, optionally accept AI-generated suggestions, and then reconcile their codes through system-supported discussion and merging~\cite{gao2024collabcoder}. This design improved efficiency while preserving reliability, as the AI handled pattern detection and summarization while human analysts retained interpretive authority. Together, these studies show that LLMs can augment qualitative coding when embedded in workflows that deliberately preserve reflexivity and analytic rigour.

In sum, existing tools often focus narrowly on code generation or treat the LLM as a black-box assistant, leaving reflexivity, concept mapping, and iterative theme development to the analyst alone. Few systems scaffold these activities from initial concept definition through code refinement and theme synthesis while maintaining researcher control. Our work contributes to this space by designing an interactive system that embeds LLM suggestions within a transparent, stepwise process: researchers define concepts, review and edit codebooks, cluster codes, and generate themes with AI assistance, but retain authority over every interpretive decision. By combining automation with human--machine dialogue, we aim to make large-scale qualitative analysis more tractable while sustaining the reflexive rigour emphasized in prior work~\cite{Long2024AIWorkflowUtility,gao2024collabcoder,xu2023aihci}.

\subsection{Trust and Adoption of Automated QDA Tools}

Trust in automated qualitative data analysis depends on researchers’ confidence that computational tools operate transparently, fairly, and reliably. Large language models are trained on vast internet corpora, and their internal mechanisms are opaque; consequently, they may reproduce dominant‑culture perspectives and misinterpret underrepresented voices~\cite{Burrell2016,chen2024dataljustice,savelka2023biasqualanalysis}. When the analytic lens involves subtle cultural differences or sensitive topics, analysts worry that an AI might reinforce biases or overlook context, undermining confidence in machine‑generated themes. This unease is compounded by the opaqueness of model predictions—researchers often cannot see why an LLM produced a specific code, making it difficult to judge whether the result reflects the data or the model’s priors.

Empirical studies of social scientists illustrate a persistent tension: qualitative researchers appreciate the potential of LLMs to handle scale, yet they question the interpretability and trustworthiness of machine outputs~\cite{Feuston2021,jiang2021serendipity}. Many fear that subtle but important content will be missed or flattened, and that algorithmic bias could distort their findings. Practical barriers also play a role: numerous qualitative researchers lack programming expertise, leading them to default to manual methods ---spreadsheets, memos or post‑it notes --- because these are familiar and under their control~\cite{Kaur2024}. In other words, researchers often prefer intensive but trustworthy manual coding to an opaque tool whose outputs they cannot fully trust~\cite{jiang2021serendipity,BraunANDClarke2019,Feuston2021}.

A consistent finding across recent evaluations is that LLMs are most effective as assistants rather than autonomous analysts~\cite{dai2023llminloop,deiner2024inductivetheme,tai2024aidanalysis}. Human oversight remains essential to ensure coding quality, contextual sensitivity and trustworthiness~\cite{chen2024dataljustice,jiang2021serendipity,kapania2025llmhciethics,savelka2023biasqualanalysis,hope2025llmqualitativeuses}. For example, \citet{dai2023llminloop} report that while an LLM could suggest reasonable open codes from raw text, human reviewers had to merge redundancies, correct misinterpretations and supply missing context. Similarly, \citet{tai2024aidanalysis} found that GPT‑4 could classify text into high‑level categories with high recall, yet it sometimes confidently produced incorrect justifications that only a human could catch. \citet{deiner2024inductivetheme} judged LLM‑generated themes as often plausible but noted that their depth and specificity seldom matched human interpretations, with the model tending toward broad or generic themes. These outcomes underscore that LLMs can accelerate coding and pattern detection, but they do not replace human interpretive judgment.

Beyond interpretability, privacy and data‑security concerns hamper adoption. Many commercial LLM services require uploading data to cloud servers, which creates ethical and legal risks when handling confidential interviews or personal histories. \citet{Perron2024} highlight that proprietary cloud‑based models “would almost certainly violate confidentiality” because whatever data are submitted can be stored, accessed or used for additional training without explicit consent, and providers rarely guarantee secure deletion. In response, researchers have begun exploring local or self‑hosted LLMs that process data entirely on users’ machines. For instance, a recent study in social work found that \emph{local} LLMs enabled analysts to classify and extract information from sensitive child‑welfare reports without transmitting any information to external servers. Such local models address confidentiality concerns while delivering performance comparable to proprietary systems, suggesting a promising path forward.

Despite growing interest, adoption of AI‑assisted qualitative analysis remains limited. Existing tools seldom support the transparency, reflexivity and data privacy that qualitative researchers expect, leaving scholars to favour manual coding over opaque automated systems~\cite{jiang2021serendipity,Feuston2021}. Our work addresses this gap by designing a locally deployable system that embeds LLM suggestions within a transparent, stepwise workflow. Analysts can inspect, edit and contextualize model outputs at each stage, maintaining control over their data and mitigating bias, misinterpretation and confidentiality risks. By foregrounding researcher agency and reflexive rigour, we hope to increase confidence in—and adoption of—automated QDA support.

\begin{figure}[tb!]
  \centering
  \includegraphics[width=\linewidth,keepaspectratio]{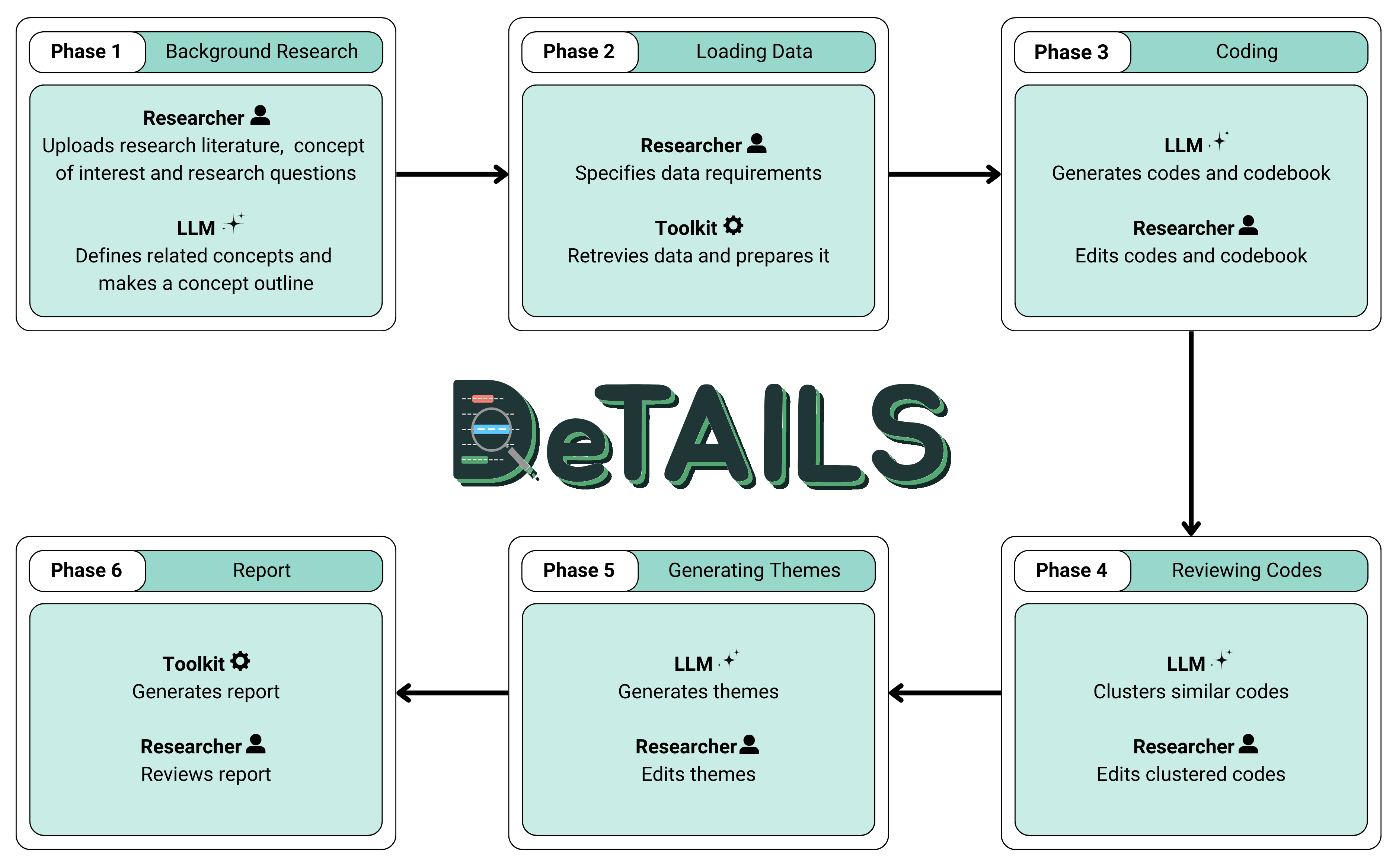}
  \caption{The DeTAILS workflow comprises six phases inspired by Braun and Clarke's Reflexive Thematic Analysis. Each phase requires collaboration between the researcher and LLM.}
  \label{tableaux-workflow}
\end{figure}

\section{DeTAILS: Deep Thematic Analysis with Iterative LLM Support}

DeTAILS is a self-contained, cross-platform desktop application that supports thematic analysis of Reddit data. Its design is inspired by Braun and Clarke’s six-phase reflexive TA framework~\cite{BraunANDClarke2021}, but intentionally adapts it into a semi-structured workflow mediated by AI support. In particular, DeTAILS uses a \emph{codebook as a boundary object} between the researcher and the LLM. While reflexive TA resists fixed codebooks, in practice a provisional, editable codebook provided a shared structure for surfacing, editing, and contesting LLM suggestions. This design choice reflects strategies often associated with codebook approaches~\cite{Boyatzis1998,roberts2019attempting}, while aiming to retain the flexibility and reflexivity central to RTA.

Researchers begin by loading data and generating an initial codebook from a subset of the dataset through human--LLM collaboration. This mirrors established TA practices where an initial subset guides broader coding~\cite{Rotolo2023}, while making the boundary between human and AI contributions explicit. DeTAILS then applies the evolving codebook to the remaining dataset and clusters codes into suggested themes, but enables researchers to iteratively reflect on, refine, and build themes using those codes.

Importantly, researchers may revisit and change any previous phase of their work, and DeTAILS automatically propagates those changes forward. For instance, a researcher may add or remove codes, redefine them, or change themes and associated codes at any point---and DeTAILS will automatically update the corresponding applications to the corpus and downstream themes. This workflow is illustrated in \autoref{tableaux-workflow}.

DeTAILS comprises a graphical front end and an analysis back end. The front end, implemented in Electron with a React UI, provides interactive interfaces for codebook creation, clustering, and theme refinement. The back end integrates a Python analysis engine with a local vector database (ChromaDB) and an LLM runtime (Ollama) to support both remote APIs and local models. A modular orchestration layer built on LangChain enables interchangeable use of four providers (OpenAI, Google Vertex AI, Google AI Studio, and local Ollama models). Each phase of the workflow is guided by structured prompts that define LLM behaviour and output format. This modular design allowed us to balance transparency, replicability, and flexibility across different model providers. Full source code is available on GitHub.

To illustrate the workflow, we now demonstrate how DeTAILS can be used to analyze posts from \textit{r/UXResearch}. This example mirrors one conducted by study participants and highlights how the toolkit scaffolds key phases of thematic analysis at scale.

\begin{figure}[tbp]
\centering
\begin{subfigure}{0.48\textwidth}
  \centering
  \includegraphics[width=\linewidth,height=0.3\textheight,keepaspectratio]{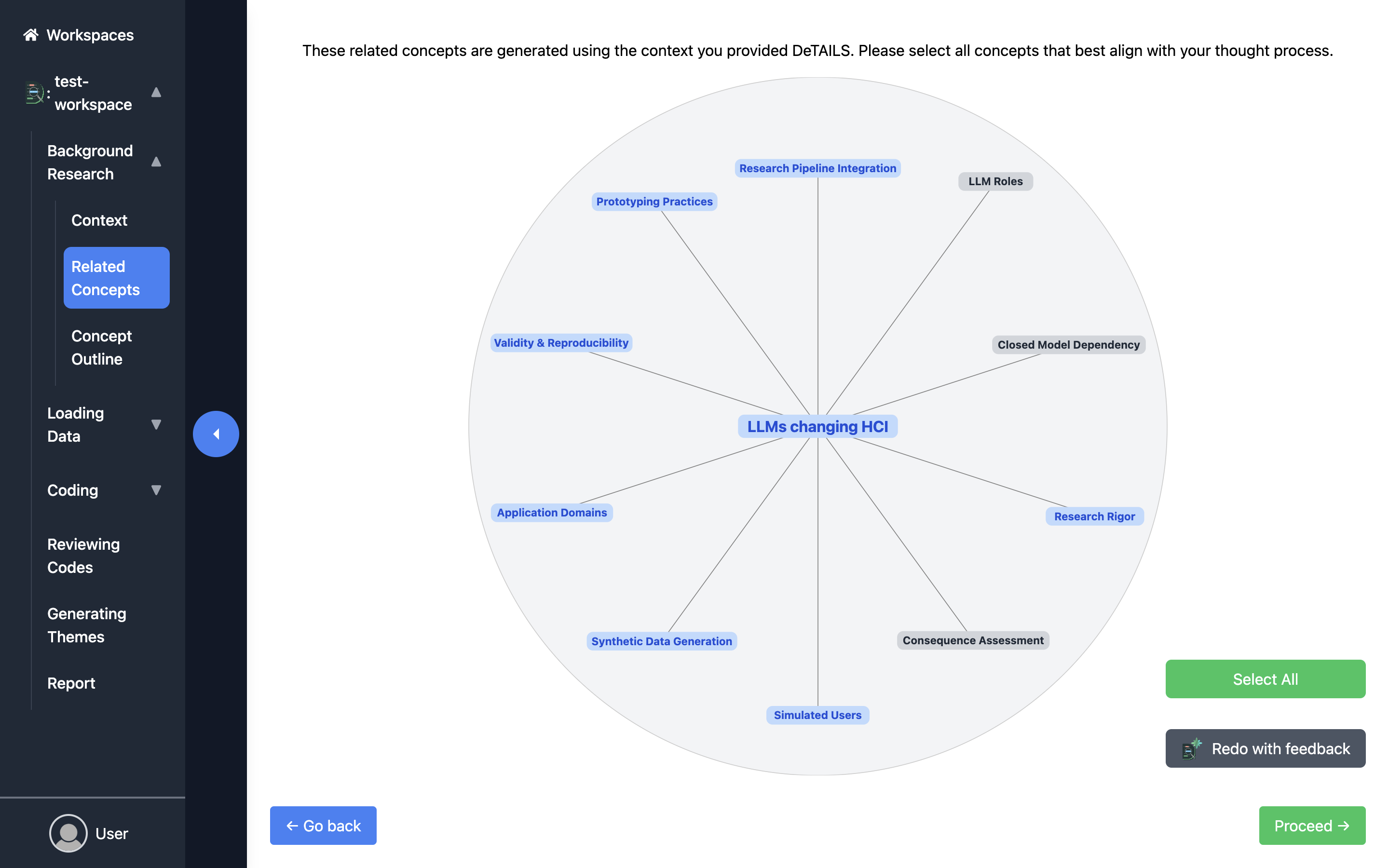}
  \caption{Background Research: Researchers provide context (e.g., research questions and related documents), and DeTAILS suggests related concepts via an interactive radial map to guide analysis.}
  \label{br-related-concepts}
\end{subfigure} \hfill
\begin{subfigure}{0.48\textwidth}
  \centering
  \includegraphics[width=\linewidth,height=0.3\textheight,keepaspectratio]{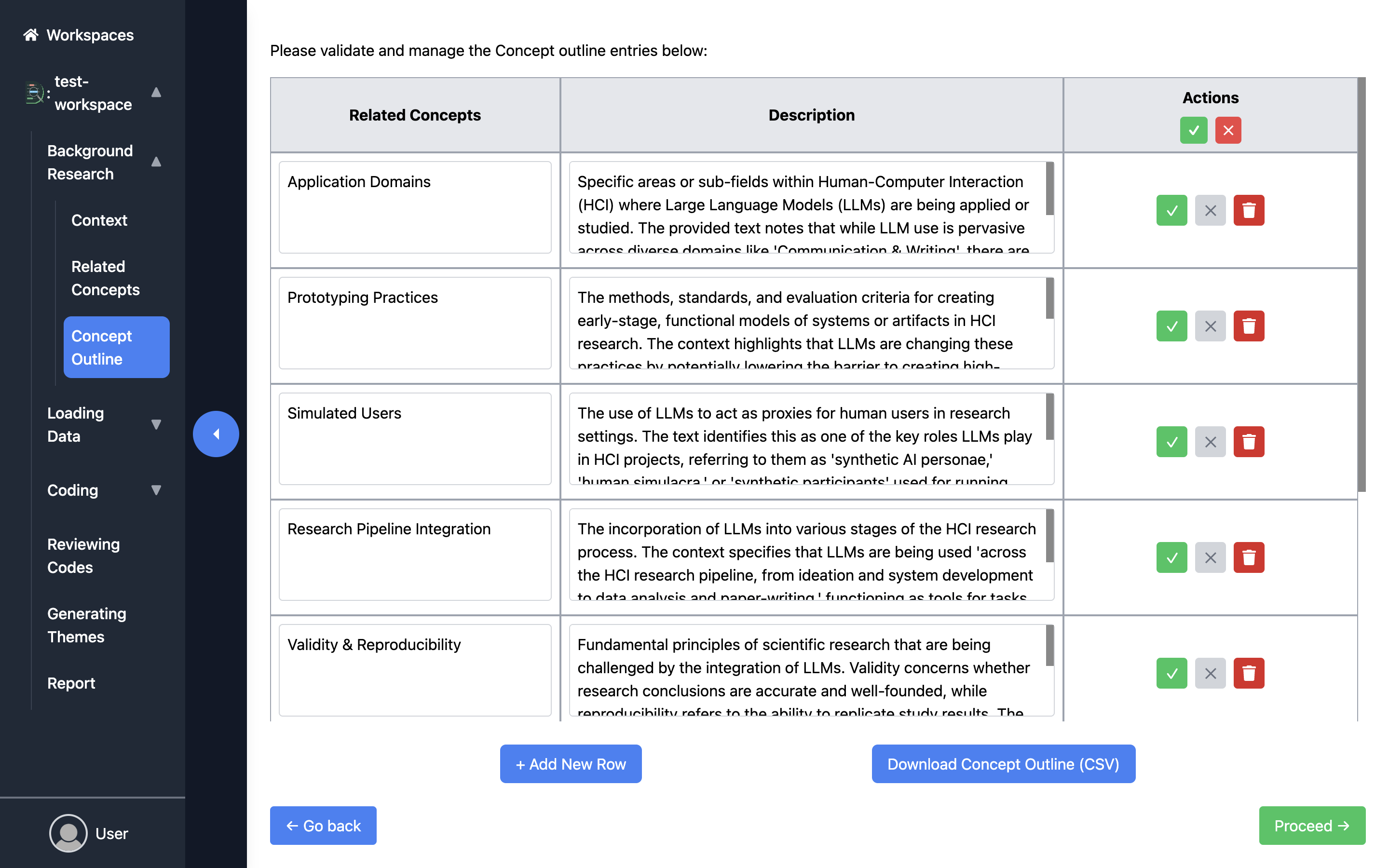}
  \caption{Concept Outline: Researchers refine and finalize a table of key concepts and definitions, which form the foundation for generating an initial codebook in later phases.}
  \label{br-concept-outline}
\end{subfigure}
\caption{Background Research: DeTAILS supports early-stage analysis by surfacing related concepts and building a structured concept outline, providing a grounded starting point for subsequent coding.}
\label{fig:background-workflow}
\end{figure}

\subsection{Phase 1: Background Research} 

Qualitative researchers begin a thematic analysis by thoroughly understanding existing research and the broader context of their research topic~\cite{Guest2012, Nowell2017}. DeTAILS embeds this process in its workflow through stored contextual information --- such as research questions and relevant literature --- provided by the researcher. Researchers provide this information by uploading related documents, like relevant research papers, and through structured and free-form text entry. The information is then processed and indexed into a ChromaDB vector database, facilitating Retrieval-Augmented Generation (RAG) so that later LLM responses are grounded in theories from the researcher’s sources. 

DeTAILs then suggests a set of related concepts based on the provided contextual information. These concepts are used as inputs to develop an initial codebook, but also provide the researcher an opportunity to reflect on the direction their analysis is taking, and to fine-tune. They are developed through a two-phase process: first researchers select concept labels~(\autoref{br-related-concepts}) from a radial map, second DeTAILS generates an editable `Concept Outline' list~(\autoref{br-concept-outline}) that includes their definitions. Once the researcher approves of these concepts, DeTAILS uses them to generate an initial codebook for anlaysis.

For instance, in the context of an analysis of ``LLMs Changing HCI'' in discussions on \textit{r/UXResearch}, the initial set of related concepts based on related literature and research questions might include concepts like `Simulated Users', `Synthetic Data Generation', or `Validity \& Rigour'. In this initial pass, the researcher might choose to exclude concepts that fall outside of their research interests like `Closed Model Dependency'. DeTAILS then generates a concept outline with more in-depth descriptions for each concept of interest.

\subsection{Phase 2: Loading Data} 

DeTAILS can load subreddit data via academic torrent dumps ~\cite{alldata, 40kdata} as well as text files from local disk. It has built-in support to load torrent files via a user's native Transmission client. It can download the files, decompress them, and filter based on the subreddit of interest. Since monthly Reddit dumps are mostly made up of data that is not of interest to the researcher, DeTAILS also provides a means to work with local, pre-filtered NDJSON files. Once loaded, the dataset can be filtered by adjusting date ranges, removing empty posts, or by keyword search. DeTAILS can similarly load data from locally provided plain text files, for instance those collected during interviews. While not supported at this time, we suggest that DeTAILS can be easily extended to work with other data sources. For instance, Reddit's upcoming API, or other social media platforms like Threads.

\begin{figure}[tbp]
\centering
\begin{subfigure}{0.48\textwidth}
  \centering
  \includegraphics[width=\linewidth,height=0.3\textheight,keepaspectratio]{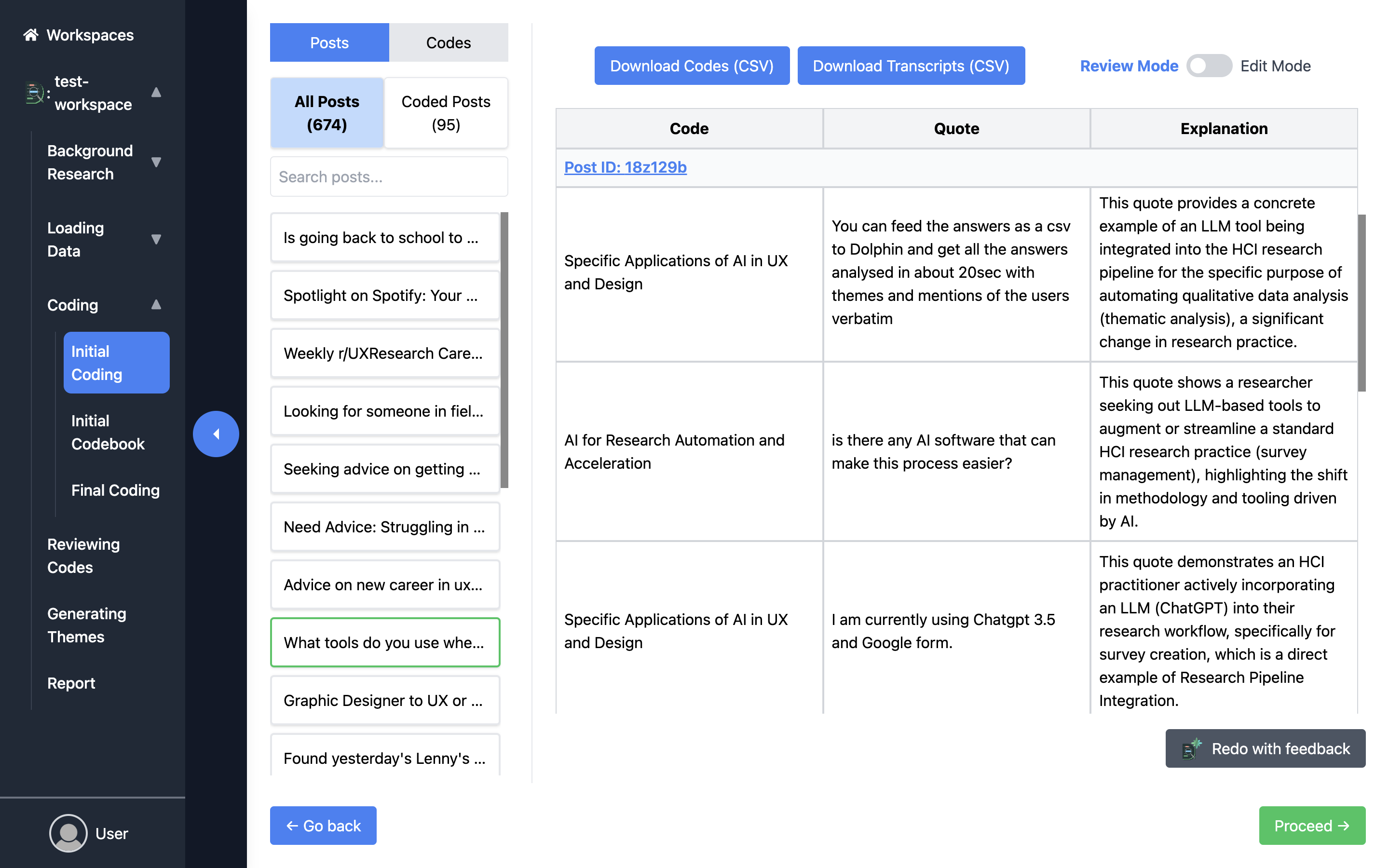}
  \caption{Initial Coding: DeTAILS applies LLM-generated codes to a subset of the data, providing a first-pass categorization.}
  \label{initial-coding}
\end{subfigure}\hfill
\begin{subfigure}{0.48\textwidth}
  \centering
  \includegraphics[width=\linewidth,height=0.3\textheight,keepaspectratio]{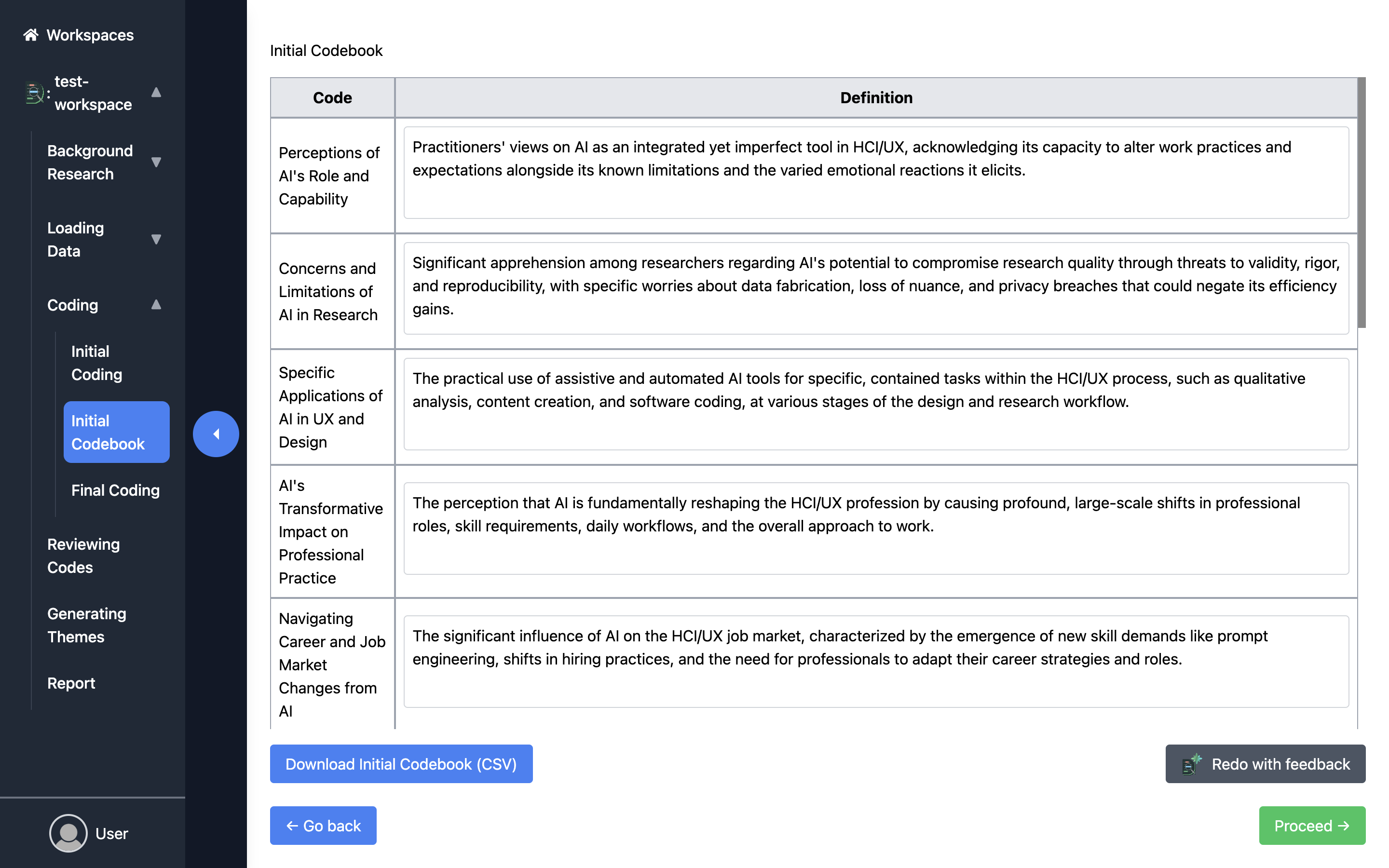}
  \caption{Initial Codebook: Researchers review and refine the LLM-suggested codebook before applying it to the full dataset.}
  \label{initial-codebook}
\end{subfigure}
\caption{Coding Phase: LLMs generate preliminary codes, which are then refined by researchers into an initial codebook to guide analysis across the corpus.}
\label{fig:coding}
\end{figure}

\subsection{Phase 3: Coding} 

In this Phase the researcher and DeTAILS work together to collaboratively define a codebook through three activities:

DeTAILS first uses the provided background information and research questions to create codes for a split of the data to create an initial codebook (\autoref{initial-coding}). The researcher can then view these codes and their application to the data through example posts and quotes, and revise them as they wish. The use of LLMs to generate this initial codebook is a key difference from previous work that relied on existing codes and examplar data for automated coding. Instead, we use these LLM-generated codes to bootstrap the rest of the coding process.

Second, the researcher can then manually edit this codebook directly (\autoref{initial-codebook}) by adjusting the code labels and definitions. DeTAILS provides the codebook in a plain text format, and the researcher can directly edit the codes by modifying their descriptive text. 

Third, the researcher can prompt DeTAILS to apply the revised codebook to the remainder of the corpus. We refer to this work as \emph{global coding}. As in the initial coding stage, DeTAILS provides a tabular interface through which the researcher can again view how the codes have been applied through a list of codes and their associated text. Researchers can also view each post individually and edit existing codes or add new codes to it.

Importantly, DeTAILS ensures the correctness of LLM-generated outputs by verifying each quote and code during each of these steps. After each LLM call DeTAILS cross-checks each quoted text segment against the original Reddit post to confirm that it appears verbatim. Any quote that cannot be found in the source text is discarded as a likely hallucination. 

% Row 3
\begin{figure}[tbp]
\centering
\begin{subfigure}{0.48\textwidth}
  \centering
  \includegraphics[width=\linewidth,height=0.3\textheight,keepaspectratio]{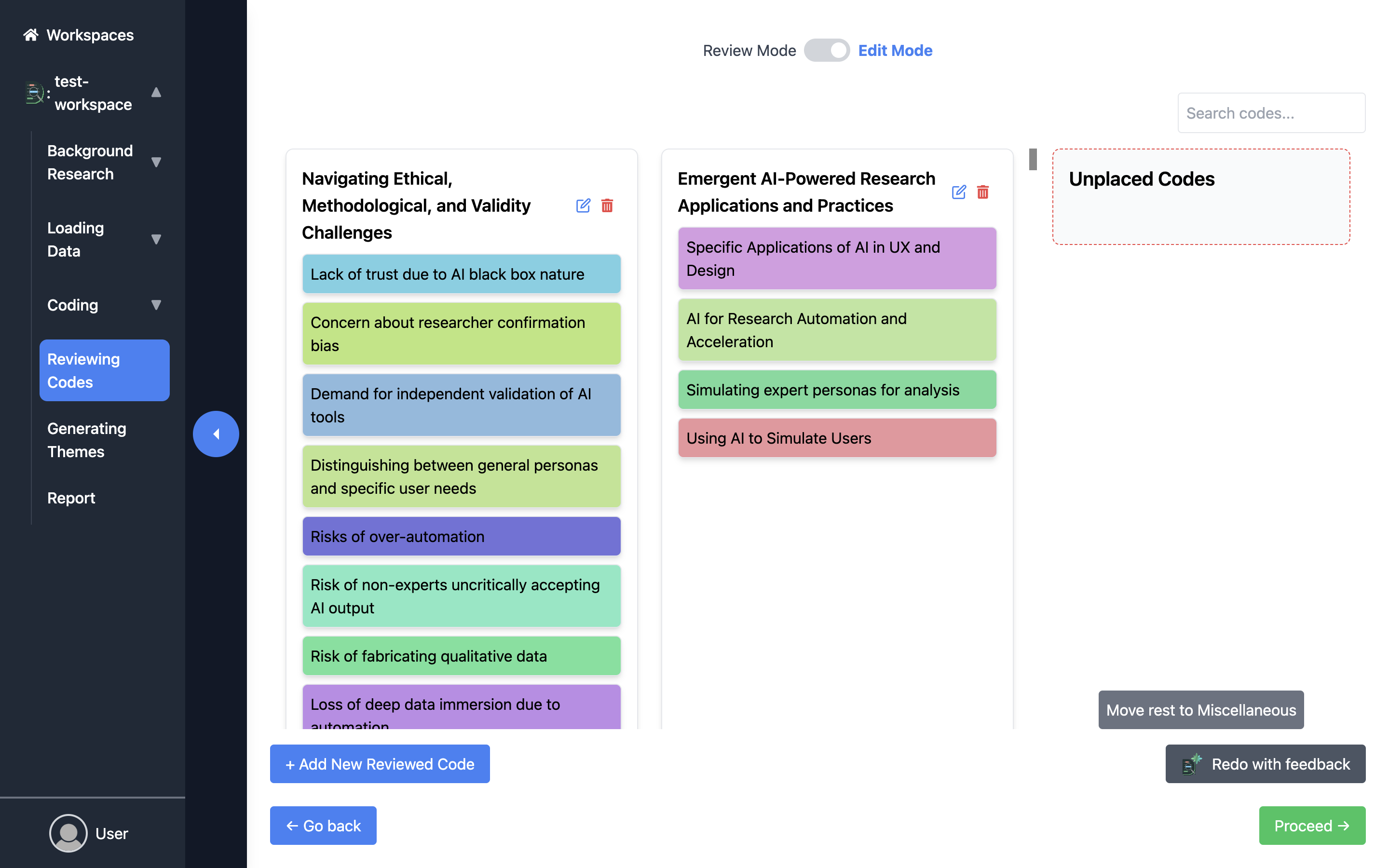}
  \caption{Reviewing Codes: DeTAILS clusters semantically similar codes into higher-level categories. Researchers can refine these clusters by editing definitions, merging, or reorganizing codes, with changes applied across the corpus.}
  \label{reviewing-codes}
\end{subfigure}\hfill
\begin{subfigure}{0.48\textwidth}
  \centering
  \includegraphics[width=\linewidth,height=0.3\textheight,keepaspectratio]{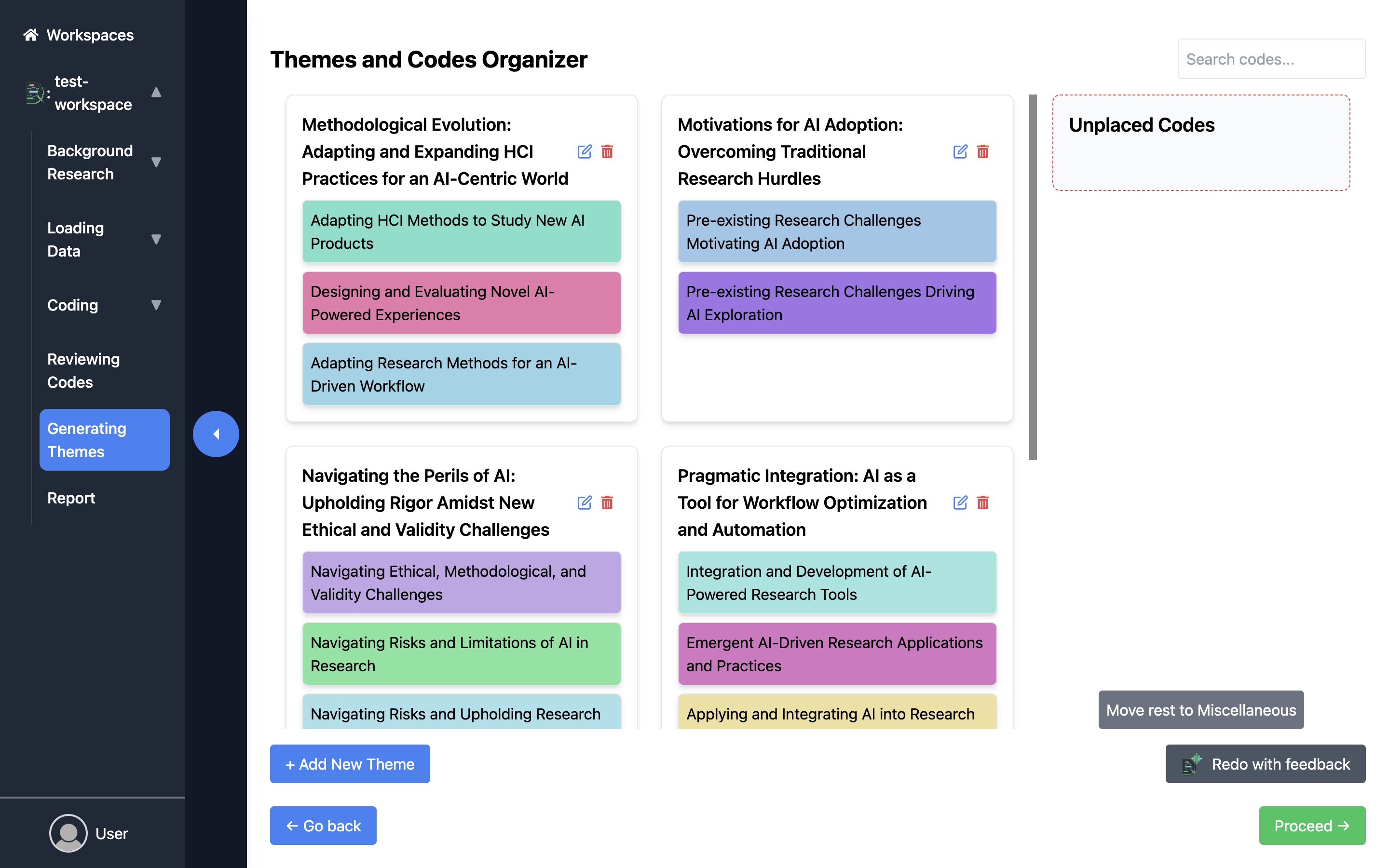}
  \caption{Generating Themes: DeTAILS proposes overarching themes from code clusters, presented in an interactive bucket interface. Researchers can adjust, merge, or create new themes to reflect their analytic lens.}
  \label{generating-themes}
\end{subfigure}
\caption{Phases 4 and 5: Reviewing Codes and Generating Themes. These phases support higher-order interpretation, allowing researchers to refine code clusters and develop coherent, researcher-driven themes.}
\label{fig:coding-and-reviewing}
\end{figure}

\subsection{Phase 4: Reviewing Codes} 

In this stage, all codes and their corresponding explanations are batched and sent to the LLM for clustering. The model groups semantically similar codes and assigns each cluster a consolidated `reviewed code'. These clusters are then visually presented via buckets, illustrating their relationships~(\autoref{reviewing-codes}). Researchers can edit the visualized clusters by dragging them between blocks or create, rename, and delete buckets. They can also fine-tune the clusters through a ‘Redo with feedback’ function that uses a text-based prompt for the LLM. Once the reviewed code groupings are finalized, an alternate view displays these codes alongside their associated quotes and explanations, maintaining a clear link to the source data. 

\subsection{Phase 5: Generating Themes} 

The final analytical step involves generating overarching themes from the reviewed codes. DeTAILS clusters all codes into a set of potential themes for the corpus. Similar to the Reviewing Codes phase~(\autoref{reviewing-codes}), these suggested themes are presented via an interactive `bucket' interface, where each theme contains a mutually exclusive set of affiliated codes~(\autoref{generating-themes}). The researcher can drag codes between themes, or create new themes, merge or rename them. DeTAILS also provides an LLM-driven refinement option, ‘Redo with feedback’ function, where researchers can explain through a text-based prompt how they would like the themes to change.

\subsection{Phase 6: Report} 

Once themes are finalized, DeTAILS assists in compiling and exporting the findings. The report interface provides options to organize the report, for instance, by `Theme \& Code' or `Post by Post'. Unlike most current tools that limit data downloads to their proprietary formats, DeTAILS enables exports of the final, structured report in the widely compatible CSV format, suitable for sharing, further analysis in other tools, or direct incorporation into publications. Although the final write‑up of the analysis still remains a manual process, DeTAILS generates a structured report --- segmenting each result into Code-Quote-Explanation sections --- that researchers can quickly review.

\section{Methodology}

We conducted a mixed-methods study to examine how qualitative researchers engaged with DeTAILS when applying thematic analysis to their own data. Our goal was to understand not only the tool’s usability and functionality, but also how it supported (or constrained) researcher agency, reflexivity, and analytic practice. Participants provided informed consent, and the study received ethics clearance from our institution’s Research Ethics Board.
%the University of Waterloo Research Ethics Board (REB \#47221).

\subsection{Participants and Recruitment}
We recruited 18 qualitative researchers with varying levels of experience and disciplinary backgrounds in reflexive thematic analysis (\autoref{tab:participant_data}). To capture a range of perspectives, we used purposive sampling with inclusion criteria of prior involvement in thematic analysis and qualitative coding. Participants were categorized by TA experience as novice (less than 1 year experience, n=6), proficient (1–4 years experience, n=6), or expert (more than 4 years experience, n=6). This stratification ensured feedback from both newcomers and seasoned practitioners.

Participants were recruited via direct email invitations using academic and professional networks. Prior to their session, each participant was asked to send their research questions, 2–5 representative research papers related to the questions, and the name of a subreddit they were familiar with and interested in analyzing.

\setlength{\tabcolsep}{2pt}
\renewcommand{\arraystretch}{1.35}

\begin{sidewaystable}[tbhp]
\checkoddpage
  \vspace*{6in}
  \centering
    \footnotesize
    \begin{tabularx}{\linewidth}{
        p{0.5cm}
        p{1.4cm}
        p{1.4cm}
        p{2.2cm}
        p{1.7cm}
        p{1.8cm}
        p{3cm}
        p{1.6cm}
        >{\RaggedRight\arraybackslash}p{5cm} 
    }
    \toprule
    \textbf{ID} & \textbf{Experience} & \textbf{Expertise} & \textbf{Field}
      & \textbf{Occupation} & \textbf{AI Used for TA} & \textbf{Subreddit}
      & \textbf{Date}
      & \textbf{Description} \\
    \midrule
    P1  & <1~year      & Novice     & XR Gaming               & Master’s\ Student & ChatGPT      & r/artificial            & 06/07/2008 -- \newline 31/12/2024
          & A subreddit dedicated to news, research, and discussions on artificial intelligence and its applications.
           \\
    P2  & <1~year      & Novice     & Explainable AI          & PhD\ Candidate   & ChatGPT      & r/ChatGPTCoding         & 06/12/2022 -- \newline 31/12/2024
          & Community for ChatGPT code samples, usage tips, and creative AI‑driven demos.
           \\
    P3  & 1--2~years   & Proficient & NLP Classification      & Jr.\ Consultant  & –      & r/WeAreTheMusicMakers   & 09/09/2008 -- \newline 31/12/2024
          & Music production, composition techniques, and audio engineering discussions.
          \\
    P4  & 2--3~years   & Proficient & {Human--AI \newline Collaboration} & PhD\ Candidate  & Atlas.TI \newline (AI features)        & r/humanresources        & 29/05/2008 - \newline 31/12/2024
          & HR best practices, talent management insights, and collaboration strategies.
          \\
    P5  & 3--4~years   & Proficient & Secure HCI              & Master’s\ Student   & GPT‑4       & r/Chatbots          & 18/01/2009 -- \newline 31/12/2024
          & Chatbot development, conversational AI frameworks, and real‑world use cases.
           \\
    P6  & 5--6~years   & Expert     & Social Stigma           & PhD\ Candidate   & NVivo \newline (AI features)         & r/beyondthebump       & 29/04/2012 -- \newline 31/12/2024
          & Support for new parents navigating postpartum and early parenting challenges.
           \\
    P7  & <1~year      & Novice     & NLP Classification      & Post‑doc\newline Researcher & Multiple LLMs   & r/ArtificialInteligence & 06/05/2016 -- \newline 31/12/2024
          & AI research, ethics debates, and industry application discussions.
          \\
    P8  & <1~year      & Novice     & Human--Robot \newline Interaction & Post‑doc\newline Researcher & Multiple LLMs  & r/CaregiverSupport      & 30/03/2013 -- \newline 31/12/2024
          & Advice and emotional support for family and professional caregivers.
           \\
    P9  & <1~year      & Novice     & Youth Smoking           & PhD\ Candidate & ChatGPT         & r/Canadian\_ecigarette   & 23/10/2013 -- \newline 31/12/2024
          & Vaping products, health impacts, and regulation updates in Canada.
          \\
    P10 & <1~year      & Novice     & Healthcare Trust        & Undergraduate\newline Student  & –      & r/AskBalkans            & 05/03/2019 -- \newline 31/12/2024
          & Q\&A on Balkan culture, travel tips, and current affairs.
         \\
    P11 & 4--5~years   & Expert     & Accessible Gaming       & PhD\ Candidate   & –        & r/disability            & 12/03/2008 -- \newline 31/12/2024
          & News, resources, and perspectives for individuals with disabilities.
         \\
    P12 & 4--5~years   & Expert     & Data Integration        & PhD\ Candidate    & –        & r/bigdata               & 05/01/2011 -- \newline 31/12/2024
          & Big data storage, processing pipelines, and predictive analytics.
        \\
    P13 & 3--4~years   & Proficient & Health Systems          & PhD\ Candidate     & –          & r/ImmigrationCanada     & 10/02/2013 -- \newline 31/12/2024
          & Immigration advice, visa procedures, and settlement experiences.
      \\
    P14 & 1--2~years   & Proficient & Haptic UX               & Product\ Analyst       & ChatGPT      & r/userexperience        & 19/06/2018 -- \newline 30/12/2024
          & UX design methods, usability testing, and prototyping best practices.
      \\
    P15 & 12--13~years & Expert     & Public Health           & PhD\ Candidate    & –        & r/blackladies           & 16/11/2012 -- \newline 31/12/2024
          & Safe space for Black women to share, support, and celebrate achievements.
        \\
    P16 & 3--4~years   & Proficient & VR Analytics            & PhD\ Candidate    & ChatGPT        & r/virtualreality        & 28/08/2011 -- \newline 31/12/2024
          & VR hardware, software platforms, and immersive design discussions.
        \\
    P17 & 10--11~years & Expert     & Dementia Technology     & PhD\ Candidate      & ChatGPT        & r/dementia              & 10/01/2010 -- \newline 31/12/2024
          & Resources and support for those affected by dementia and their caregivers.
      \\
    P18 & 11--12~years & Expert     & Human Factors           & Assistant\newline Professor    & –   & r/smartwatch            & 21/01/2013 -- \newline 31/12/2024
          & Smartwatch tech, app development, and wearable computing trends.
       \\
    \bottomrule
    \end{tabularx}

  \caption[Study participants demographics and subreddit]{Demographics of Study Participants: Years of Experience, TA Expertise Level, Field of Specialization, Current Occupations, AI used for TA, Subreddit details: Name, Start Date - End Date, Description}
    \label{tab:participant_data}
\end{sidewaystable}

\subsection{Materials and Apparatus}

We focused on Reddit data given its heterogeneity and scale~\cite{proferes2021}, which has led to its use in numerous HCI studies (e.g.,~\cite{Rotolo2023,chandrasekharan2018,jhaver2019,li2021}). Working with Reddit data allowed us to evaluate DeTAILS on large-scale, noisy real-world text, while also enabling each participant to explore content relevant to their interests. We then pre-loaded the participant’s chosen subreddit data into DeTAILS. For each participant, we downloaded all posts and comments from the last full month \cite{40kdata} and loaded a new workspace for participants to use for their study session.

Sessions were conducted using a MacBook Air with an Apple M1 chip (8‑core CPU, 7‑core GPU) and 16 GB RAM. This machine ran DeTAILS with the gemini-2.5-pro-preview-03-25 LLM via the Google Vertex AI API. Participants who attended in person used this laptop directly. For remote participants, we utilized Microsoft Teams’ screen-sharing and remote control functionality: the author shared their screen running DeTAILS, and the participant was granted control to operate the interface virtually. This setup allowed online participants to fully interact with the toolkit (e.g., clicking buttons, highlighting text, editing code labels) as if they were seated at the machine. Audio and video were streamed via Teams for communication. Two participants opted for in-person sessions in our lab, while the remaining 16 participated remotely via Teams. 

\subsection{Study Procedure}
Each study session lasted between 2 and 2.25 hours and followed a structured sequence: an initial semi-structured interview, a guided walkthrough of DeTAILS, an interactive analysis task using DeTAILS, a follow-up exploration of large-scale results, and post-study questionnaires. 

The central activity was having participants use DeTAILS to conduct thematic analysis on 30 discussion post transcripts from a subreddit of their choice (each transcript consisting of an original post and its comments). We selected a subset of 30 posts to ensure that interactive analysis was feasible within the session time while still providing sufficient material to generate meaningful codes and themes. Participants were asked to approach this analysis as if it were part of their own research. The task was structured according to the toolkit’s workflow, with participants in control and the researcher observing, supporting, and prompting reflection when needed.  

After completing the 30-post analysis, participants were introduced to pre-computed results for the full month of subreddit data (December 2024) that had been prepared in advance. They were able to browse and inspect outputs from each phase at this larger scale. This stage allowed participants to reflect on how the workflow operated across two scales, and to provide additional feedback on features, usability, and desired improvements. Participants could ask clarifying questions about the tool throughout.  

The first author designed the study protocol in consultation with the third author and conducted all participant interviews. 

\subsection{Data Collection and Analysis}
We collected both quantitative and qualitative data. Quantitative data included system logs of participant interactions (e.g., time per phase, edits, suggestion usage) and three post-study surveys (NASA-TLX, AttrakDiff, Perceived Usefulness). Statistical analyses were conducted by the first author, with consultation from the third author. We computed weighted precision, recall, and F1 scores using cosine similarity across Related Concepts, Concept Outline, Initial Coding, and Global Coding; semantic similarity for Initial Codebook definitions; and macro F1 metrics for Reviewing Codes and Generating Themes. Time across study phases was analysed using repeated-measures ANOVA with Mauchly’s test for sphericity; Greenhouse-Geisser corrections or Friedman tests were used when assumptions were violated. Post-hoc Wilcoxon signed-rank tests (Bonferroni-corrected) followed significant effects. For group comparisons (Novices, Proficients, Experts), assumption checks (Shapiro-Wilk, Levene’s) guided use of one-way ANOVA, Welch’s ANOVA, or Kruskal-Wallis tests. Pairwise comparisons employed t-tests, Welch’s t-tests, or Mann-Whitney U tests with Bonferroni-adjusted $\alpha = 0.0167$.

Qualitative data came from the post-session semi-structured interviews and conversations while the participant was using the system, which were transcribed verbatim. The second author, an expert in qualitative analysis, led the thematic analysis with input from the other authors. This analysis was conducted manually rather than in DeTAILS, but emulated the codebook-style workflow that the system itself supports. Codes and themes were iteratively refined over three rounds, a process resembling the recursive engagement of reflexive TA but operationalised in a more structured, codebook-oriented manner. This approach enabled us to capture patterns in participant reasoning, experience, and feedback across the phases of the workflow.

\section{Quantitative Results}

We report results on (1) alignment between DeTAILS-generated outputs and participants’ refinements (F1 scores), (2) time spent across workflow phases, and (3) user-reported workload (NASA-TLX), perceived usefulness, and pragmatic/hedonic qualities (AttrakDiff).

\subsection{F1 Scores}
\begin{figure*}[tbp]
\centering
\begin{tikzpicture}
\begin{groupplot}[
    group style={
      group size=4 by 1,
      horizontal sep=.75cm,
    },
    IDEA box,               
    ymin=0, ymax=1,
    %width=0.27\textwidth,   
    height=0.35\textwidth,
    xtick=\empty
]

% Phase 1 (double width, with y-axis)
\nextgroupplot[
    title={\shortstack{Phase 1\\Background Research}},
    %width=0.4\textwidth,
    ytick={0,0.2,0.4,0.6,0.8,1},        % positions for horizontal lines
    ymajorgrids=true,                   % enable grid
    xtick={1,2},
    xticklabels={Related Concepts, Concept Outline},
    x tick label style={align=center},
    ylabel={F1 Score},
    x=2.25cm,
    boxplot/box extend=0.25
]
\addplot+ [boxplot,draw=black] table[y=Phase1, col sep=comma] {data/f1-scores.csv};
\addplot+ [boxplot,draw=black] table[y=Phase1b, col sep=comma] {data/f1-scores.csv};

% Phase 3
\nextgroupplot[
    title={\shortstack{Phase 3\\Coding}},
    ytick={0,0.2,0.4,0.6,0.8,1},        % positions for horizontal lines
    yticklabels=\empty,                 % hide labels
    ymajorgrids=true,                   % enable grid
    y axis line style={draw=none},      % hide the vertical axis line
    xtick={1,2}, xticklabels={Codebook, Global Coding},
    x=2.25cm,
    boxplot/box extend=.25
]
\addplot+ [boxplot,draw=black] table[y=Phase2, col sep=comma] {data/f1-scores.csv};
\addplot+ [boxplot,draw=black] table[y=Phase3, col sep=comma] {data/f1-scores.csv};

% Phase 4
\nextgroupplot[
    title={\shortstack{Phase 4\\\vphantom{Background Research}}},
    ytick={0,0.2,0.4,0.6,0.8,1},        % positions for horizontal lines
    yticklabels=\empty,                 % hide labels
    ymajorgrids=true,                   % enable grid
    y axis line style={draw=none},      % hide the vertical axis line
    xtick={1}, xticklabels={Reviewing Codes},
    x = 1.5cm,
    boxplot/box extend=0.35
]
\addplot+ [boxplot,draw=black] table[y=Phase4, col sep=comma] {data/f1-scores.csv};

% Phase 5
\nextgroupplot[
    title={\shortstack{Phase 5\\\vphantom{Background Research}}},
    ytick={0,0.2,0.4,0.6,0.8,1},        % positions for horizontal lines
    yticklabels=\empty,                 % hide labels
    ymajorgrids=true,                   % enable grid
    y axis line style={draw=none},      % hide the vertical axis line
    xtick={1}, xticklabels={Generating Themes},
    x = 1.5cm,
    boxplot/box extend=0.35
]
\addplot+ [boxplot,draw=black] table[y=Phase5, col sep=comma] {data/f1-scores.csv};

\end{groupplot}
\end{tikzpicture}
\caption{Box and whisker plot of F1 Scores across each phase of reflexive thematic analysis. Background Research and Coding are measured by a weighted F1, whereas Reviewing Codes and Generating Themes use a Macro F1.}
\label{fig:f1-all}
\end{figure*}
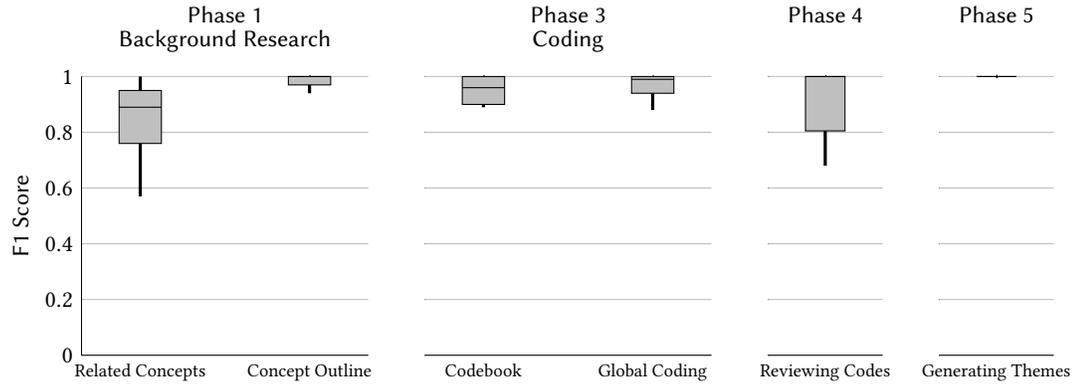

F1 scores increased consistently across phases (\autoref{fig:f1-all}). In Background Research, Related Concepts suggestions achieved a mean weighted F1 of 0.86 (SD = 0.13), improving to 0.98 (SD = 0.03) at Concept Outline. In the Coding phase, Initial Coding averaged 0.90 (SD = 0.17), rising to 0.97 (SD = 0.04) in Global Coding. Reviewing Codes produced macro F1 = 0.90 (SD = 0.20), and Generating Themes reached complete agreement (F1 = 1.00).

No significant expertise differences emerged. Related Concepts satisfied parametric assumptions, yielding a nonsignificant ANOVA ($F_{2,15} = 0.22$, $p = 0.8032$). Other phases violated normality but not homoscedasticity; Kruskal–Wallis tests were nonsignificant ($H_{2} = 1.95$–$5.82$, $p = 0.0544$–$0.7664$). Generating Themes was excluded due to zero variance.

% \begin{figure}[tb!]
%   \centering
%   \includegraphics[width=\textwidth]{F1-all.png}
%   \caption{F1 Scores: Background Research Phase (Related Concepts, Concept Outline), Coding Phase (Initial Coding, Global Coding), Reviewing Codes Phase, Generating Themes Phase}
%   \label{fig:f1-all}
% \end{figure}

\subsection{Time Analysis}
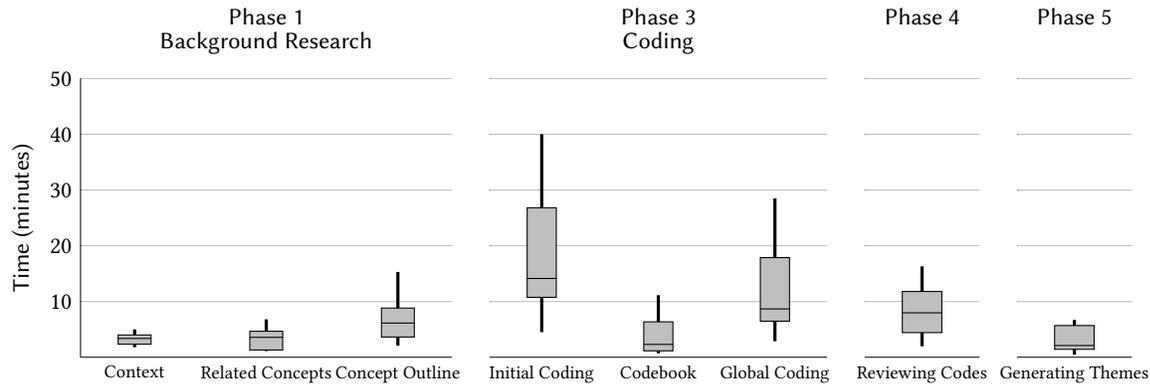
\begin{figure*}[tbp]
\centering
\begin{tikzpicture}
\begin{groupplot}[
    group style={
      group size=4 by 1,
      horizontal sep=.5cm,
    },
    IDEA box,               
    ymin=0, ymax=50,
    %width=0.27\textwidth,   
    height=0.35\textwidth,
    xtick=\empty
]

% Phase 1 (double width, with y-axis)
\nextgroupplot[
    title={\shortstack{Phase 1\\Background Research}},
    %width=0.4\textwidth,
    ytick={10,20,30,40,50},         % positions for horizontal lines
    ymajorgrids=true,                   % enable grid
    xtick={1,2,3},
    xticklabels={Context,Related Concepts, Concept Outline},
    x tick label style={align=center},
    ylabel={Time (minutes)},
    x=1.75cm,
    boxplot/box extend=0.25
]
\addplot+ [boxplot,draw=black] table[y=Context, col sep=comma] {data/time.csv};
\addplot+ [boxplot,draw=black] table[y=RelatedConcepts, col sep=comma] {data/time.csv};
\addplot+ [boxplot,draw=black] table[y=ConceptOutline, col sep=comma] {data/time.csv};

% Phase 3
\nextgroupplot[
    title={\shortstack{Phase 3\\Coding}},
    ytick={10,20,30,40,50},         % positions for horizontal lines
    yticklabels=\empty,                 % hide labels
    ymajorgrids=true,                   % enable grid
    y axis line style={draw=none},      % hide the vertical axis line
    xtick={1,2,3}, xticklabels={Initial Coding,Codebook, Global Coding},
    x=1.55cm,
    boxplot/box extend=.25
]
\addplot+ [boxplot,draw=black] table[y=InitialCoding, col sep=comma] {data/time.csv};
\addplot+ [boxplot,draw=black] table[y=InitialCodebook, col sep=comma] {data/time.csv};
\addplot+ [boxplot,draw=black] table[y=GlobalCoding, col sep=comma] {data/time.csv};

% Phase 4
\nextgroupplot[
    title={\shortstack{Phase 4\\\vphantom{Background Research}}},
    ytick={10,20,30,40,50},         % positions for horizontal lines
    yticklabels=\empty,                 % hide labels
    ymajorgrids=true,                   % enable grid
    y axis line style={draw=none},      % hide the vertical axis line
    xtick={1}, xticklabels={Reviewing Codes},
    x = 1.5cm,
    boxplot/box extend=0.35
]
\addplot+ [boxplot,draw=black] table[y=ReviewingCodes, col sep=comma] {data/time.csv};

% Phase 5
\nextgroupplot[
    title={\shortstack{Phase 5\\\vphantom{Background Research}}},
    ytick={10,20,30,40,50},         % positions for horizontal lines
    yticklabels=\empty,                 % hide labels
    ymajorgrids=true,                   % enable grid
    y axis line style={draw=none},      % hide the vertical axis line
    xtick={1}, xticklabels={Generating Themes},
    x = 1.5cm,
    boxplot/box extend=0.35
]
\addplot+ [boxplot,draw=black] table[y=GeneratingThemes, col sep=comma] {data/time.csv};

\end{groupplot}
\end{tikzpicture}
\caption{Box and whisker plot of time spent across each phase of reflexive thematic analysis. }
\label{fig:time-all}
\end{figure*}

Time analysis showed Initial Coding took significantly longer than other phases (\autoref{fig:time-all}). Mauchly’s test indicated violation of sphericity ($W = 0.197$, $p = 0.003$); applying Greenhouse–Geisser ($\varepsilon = 0.623$), repeated-measures ANOVA revealed a strong effect of Phase ($F_{2.49, 42.37} = 22.77$, $p < 0.001$, $\eta^2_p = 0.57$), corroborated by a Friedman test ($\chi^2_{4} = 44.62$, $p < 0.001$). Post-hoc Wilcoxon tests (Bonferroni-corrected) showed Initial Coding was significantly longer than Related Concepts, Concept Outline, Initial Codebook, and Global Coding ($p \leq 0.004$). Global Coding also exceeded Initial Codebook ($p = 0.004$).

By expertise, only Related Concepts differed: Welch’s ANOVA ($F_{2,9.26} = 6.51$, $p = 0.017$) showed Proficients (M = 5.37 min) spent more time than Novices (M = 2.49 min, $p = 0.004$). No other group differences were significant.

% \begin{figure}[tb!]
% \centering
% \includegraphics[width=\textwidth]{figures/time-all.png}
% \caption[Participant timings (by phase) while using DeTAILS]{Time spent per phase: Related Concepts, Concept Outline, Initial Coding, Global Coding, Reviewing Codes, Generating Themes.}
% \label{fig:time-all}
% \end{figure}

\subsection{NASA-TLX}
Participants reported low to moderate workload, with means of 39/100 (mental demand), 37/100 (effort), 29/100 (temporal demand), and notably low frustration (median 10/100). Self-rated performance was high (mean 73/100). Overall workload averaged 26.3/100 (SD = 12.4). We found no significant differences between expertise groups. One-way ANOVA, Welch’s ANOVA, or Kruskal–Wallis tests found no effects for any subscales ($p = 0.1993$–$0.8919$).

\subsection{Perceived Usefulness Scale}
Participants rated DeTAILS highly useful (M = 4.21/5), with 86\% of responses at 4 or 5. Strongest items were “accomplishes tasks more quickly” (M = 4.4/5) and “makes it easier to do my job” (M = 4.4/5). Fifteen of 18 participants rated at least four items as 4 or 5. Group comparisons revealed no significant differences. Only “Improves Performance” met parametric assumptions, yielding a nonsignificant ANOVA ($F_{2,15} = 0.17$, $p = 0.8433$). Other items tested with Kruskal–Wallis were also nonsignificant ($p = 0.1137$–$0.8679$).

\subsection{AttrakDiff}
AttrakDiff ratings were overwhelmingly positive. Pragmatic qualities scored highly (e.g., structured: M = 6.3/7, straightforward: M = 5.7/7, simple: M = 2.8/7 where 1 = simple). Hedonic ratings were similarly strong (e.g., creative: M = 5.9/7, captivating: M = 5.8/7). No significant expertise effects were observed for most items ($p = 0.1196$–$1.0000$). A trend emerged for Simple/Complicated ($\chi^2_{2} = 5.88$, $p = 0.0529$): novices (Mdn = 2.00) perceived DeTAILS as simpler than experts (Mdn = 3.00), with post-hoc tests suggesting a difference at corrected $\alpha$.

\section{Qualitative Results}
Our analysis produced five overarching themes that capture how researchers navigated their interactions with DeTAILS: asserting agency, evaluating performance, negotiating trust, experiencing partnership, and reflecting on emotional and ethical implications. Within each theme, participants described both opportunities and challenges of integrating AI into qualitative workflows.

\subsection{Asserting Researcher Agency \& Control}
Participants emphasized that DeTAILS did not replace their analytic authority but required active oversight. Two codes captured this theme. The first, \textit{Agency Through Understanding AI’s Limitations}, described how participants stressed the need for human interpretation. P7 explained, “It cannot be completely automated with an AI model. It needs some input… it is very difficult to completely automate,” underscoring that analytic work requires subjectivity and nuance. The second code, \textit{Verifying AI Output as an Act of Agency}, highlighted validation as a deliberate responsibility. P17 reflected, “I didn’t agree with the explanation… the code was fine, but the quote I felt didn’t align,” showing how participants treated verification as an assertion of responsibility. Together, these codes show that agency was reasserted through recognition of AI’s limits and critical oversight of its outputs.

\subsection{Evaluating the Tool’s Performance \& Utility}
Participants evaluated DeTAILS in terms of benefits and shortcomings, expressed across seven codes. The code \textit{Accelerating the Workflow} captured how participants valued speed, with P1 observing, “This was definitely faster… what would take hours was done in less than 5 minutes.” Yet participants also cautioned that speed did not guarantee better analysis, captured in the code \textit{Critiquing Lack of Context or Nuance}. As P10 explained, “There’s a big gap in what the LLM was able to do… the quotes were a bit less nuanced, a little less personal,” illustrating how efficiency sometimes came at the cost of depth.

A second cluster of codes focused on comparisons and data issues. The code \textit{Comparison with Existing Tools and Methods} highlighted how participants situated DeTAILS against their existing practices. P16 admitted, “I wasn’t a huge fan of the initial coding… I would rather do it myself. Once I do some percentage… then maybe the AI can add.” The code \textit{Data-Specific Challenges and Observations} emphasized how limitations in the Reddit dataset shaped analytic outcomes. P12 noted, “When people are posting recent information, the number… is so low that it might not appear as a theme.” These codes together show that the system’s performance was tied not only to its design but also to the characteristics of the data.  

The final set of codes pointed to usability and adoption. \textit{Evaluating Tool Design and Usability} emphasized smooth interactions but also highlighted areas of friction; P18 appreciated visualizations but wished features were easier to find. \textit{Future Use and Adoption} captured enthusiasm about integrating the tool into future projects, with P14 reflecting, “I enjoyed using it… I know it could have a very big impact, especially for researchers going through a lot of data.” Finally, \textit{Valuing High-Quality, Nuanced AI Contributions} described moments when participants felt the system added real analytic value. P15 noted, “The sophisticated language helped with the grouping in themes… it wasn’t just like a 1+1=2.” Together, these codes illustrate how participants weighed speed, usability, and nuance when assessing DeTAILS' overall utility.

\subsection{Trust and Verification}
Trust in DeTAILS was not automatic but had to be earned through explanation, alignment, and validation. Two codes captured this negotiation directly. \textit{Building Trust Through Agreement and Disagreement} described how alignment with expectations could strengthen confidence, as P17 shared, “…this again verifies what I’m thinking and also builds upon it.” In contrast, \textit{Distrust in Opaque Processes} pointed to skepticism when outputs felt shallow or overly accommodating. P4 remarked, “…sometimes I feel like the AI is trying to manage our expectations… so it’s like false positives.”

Three additional codes captured how participants sustained trust through ongoing oversight. \textit{Researcher Validation as a Core Task} reflected how outputs were systematically checked, with P13 describing them as “probably accurate… to the few select number of posts we were asked to review.” \textit{Trust Through Transparency and Explainability} emphasized that explanations were a foundation for cautious trust, though they did not replace human judgment. As P11 explained, “If I was using this for a study, I would spend probably a week… really checking it.” Finally, the code \textit{Valuing AI Uncertainty to Support Judgment} described how participants sometimes treated ambiguous outputs as prompts for reflection rather than flaws. Together, these codes illustrate that trust in DeTAILS was relational and conditional, anchored in researcher validation and critical engagement.

\subsection{The Human–AI Partnership}
Participants also framed their work with DeTAILS as a kind of partnership, represented across five codes. \textit{Framing the AI as a Teammate or Partner} captured how participants saw the system as collaborative, though never autonomous. P11 remarked, “If the LLM is my partner… I would still probably take the step of reading these things over.” \textit{Experiencing Dialogue and Co-Interpretation} described how outputs prompted reflection and refinement; P2 explained, “The requirement for prompting was really low… it really streamlines things,” highlighting a conversational rhythm of feedback and adjustment.

The code \textit{Positioning AI as a Second Opinion} captured how outputs were often treated as confirmatory checks, with P13 noting they were “probably accurate… to the few select number of posts we asked to review.” \textit{Recognizing the Boundaries of Partnership} emphasized that interpretive authority remained with humans, as P7 stressed, “It cannot be completely automated… it needs some input.” Finally, \textit{Negotiating Roles in Collaborative Analysis} captured how researchers actively shaped their inputs to direct the AI’s contributions, with P12 reflecting, “…when you’re looking for quality… that’s what problem I need to change the questions a little bit.” Together, these codes show that participants positioned DeTAILS as a supportive partner that could accelerate, provoke, or confirm analysis, but only within carefully defined boundaries.

\subsection{The Reflexive \& Emotional Experience}
Participants reflected on the emotional and reflexive dimensions of working with DeTAILS, expressed across five codes. The code \textit{Experiencing Surprise and Delight} captured moments when outputs exceeded expectations. P1 noted, “It brought out topics that I didn’t even think it would… thinking about the future and understanding user demographics, that’s a really good thing.” In contrast, \textit{Encountering Frustration or Skepticism} emphasized when outputs felt shallow or misleading, as P4 explained, “…sometimes I feel like the AI is trying to manage our expectations… so it’s like false positives.”

The code \textit{Reflecting on Bias and Preconceptions} highlighted how participants became aware of problematic assumptions in AI systems. P11 recalled, “I was using ChatGPT… and it literally used the R slur in one of the suggested names,” underscoring risks of bias in outputs. \textit{Emotional Responses as Drivers of Reflexivity} described how emotions shaped reflexive engagement; P10 admitted, “I feel extremely lazy and I feel like I don’t deserve to have this title of researcher, if I’m using an LLM to do my research for me,” showing how discomfort prompted reflection on researcher identity. Finally, \textit{Considering Ethical or Existential Implications} drew attention to risks around data integrity and representation. P17 warned, “If you forget to remove an identifier… I just worry about the safety of the data.” Collectively, these codes highlight that working with AI was as much affective and ethical as it was technical, prompting participants to critically situate the tool within their research practice.

Across these five themes and their associated codes, participants portrayed their engagement with DeTAILS as a complex negotiation of efficiency, trust, control, partnership, and reflexivity. They valued the system’s ability to accelerate tasks, provoke insights, and act as a supportive collaborator, while also emphasizing the need for human oversight, boundary-setting, and ethical responsibility. Their accounts suggest that integrating AI into qualitative research is not merely a technical matter but a relational and ethical practice in which analytic authority remains firmly with the researcher.

\section{Discussion}

DeTAILS was designed to explore whether LLMs could accelerate the labour of qualitative coding while maintaining researcher agency, reflexivity, and trust. Our evaluation showed substantial efficiency gains: participants completed datasets in about 33 minutes compared to the seven to eight hours typically required for manual analysis. Survey data confirmed perceptions of efficiency and usability, and log data indicated high alignment between model outputs and human refinements. At the same time, participants stressed the need for careful review and refinement, highlighting that analytic authority remained with the researcher.

Relative to prior systems such as NVivo auto-coding \cite{NVivoHelpAutocode2025} or topic-modelling approaches \cite{Chen2023, Miyaoka2023}, DeTAILS offers both efficiency and interpretive alignment. High agreement scores and positive manageability ratings underscore how stepwise verification checkpoints scaffolded trust, leading participants to describe the AI as a “junior coder” or “second opinion.” This reframes AI from a black-box generator to a collaborative partner. This also situates DeTAILS within broader debates about ‘big qual’ methods \cite{jamieson2024, abhinaya2024bigqual}, which seek to scale qualitative analysis without abandoning depth. Our results show that AI can accelerate labour while maintaining researcher engagement, extending these debates beyond descriptive approaches like topic modelling. Our contribution goes beyond scaling descriptively --- it shows how interpretive engagement can also be scaffolded at scale.

\subsection{Implications for Theory and Reflexivity}

Although DeTAILS scaffolded reflexivity, it did not fully achieve the iterative, self-critical reflexivity central to RTA \cite{BraunANDClarke2019}. Log data showed that most participant time was spent reworking initial codes rather than revisiting themes, and the system’s forward-only propagation meant refinements at later stages could not cascade backwards. This design, coupled with participants’ survey responses that they would “still read everything over,” produced a workflow resembling codebook thematic analysis \cite{Boyatzis1998, Guest2012}: reflexivity occurred through checking and correcting model outputs, rather than interrogating one’s own analytic assumptions. That participants reported high efficiency and usability while still expressing caution --- P11 estimated she would spend “a week checking” before trusting the outputs --- illustrates how quant and qual results converge on this tension.

Quantitative indicators also highlight the limits of automation. Despite high F1 agreement, participants noted that quotes were “less nuanced” (P10) and that the outputs often lacked personal tone. This mismatch shows that statistical alignment does not equate to interpretive adequacy, echoing concerns that LLMs flatten complexity \cite{chew2024aiqualresearch, savelka2023biasqualanalysis}. AttrakDiff trends hinted at a “canny mountain” effect: novices were more likely to describe the system as “human” while experts were more sceptical, suggesting risks of over-trust in polished outputs. These findings underscore that some “grunt work” of qualitative analysis—iterative reading, grappling with ambiguity—remains irreducibly human and methodologically valuable \cite{Finlay2002, Nowell2017}.

Taken together, our results suggest a division of labour. Quantitatively, DeTAILS provides strong evidence of efficiency, usability, and alignment; qualitatively, participants confirmed that the AI accelerated but did not replace analytic reasoning. This shows that LLMs may be well suited for scaffolding codebook-style workflows, while RTA’s deeper reflexivity demands design features not yet realized. Rather than seeing this as a failure, we argue it clarifies the epistemological stakes: AI can support, but not substitute, the slow and interpretive work that gives qualitative research its depth \cite{BraunANDClarke2020, Williams2024paradigm}. By explicitly delineating where automation adds value (speed, reliability, consistency) and where human labour must remain central (reflexivity, meaning-making, ethics), future tools can extend thematic analysis to larger datasets without eroding its methodological integrity.

Our findings also speak to theoretical debates about whether LLMs can meaningfully participate in qualitative inquiry. Reflexive thematic analysis (RTA) emphasizes researcher subjectivity, positionality, and iterative meaning-making as central to the analytic process \cite{BraunANDClarke2019, Finlay2002}. DeTAILS supported some of this reflexive work, but in ways that shifted its orientation: participants’ reflexivity often focused on questioning the AI rather than questioning their own assumptions. This suggests that AI tools can scaffold certain forms of reflexivity, but risk narrowing it to verification unless deliberately designed to foreground positional reflection.

We therefore conclude that the role of AI in qualitative research should not be imagined as replacing human interpretation, but as scaffolding it \cite{baumer2017comparing}. By recognizing the stages where AI support is productive, the stages where human labour remains indispensable, and the theoretical commitments that must guide design, we can move toward tools that respect the epistemological richness of qualitative inquiry while extending its reach. Ultimately, our study shows that AI can extend --- not diminish --- the epistemic value of qualitative inquiry, but only when systems are designed to centre researcher reflexivity, ethics, and interpretive authority. In doing so, DeTAILS contributes to ongoing debates about how qualitative research can adapt to scale while preserving its interpretivist commitments. 

%Acknowledging this boundary is crucial for interpretivist traditions: some analytic labour ---iterative reading, grappling with ambiguity, and holding space for participants’ voices --- cannot be automated without undermining the epistemological commitments of reflexive qualitative inquiry~\cite{Finlay2002, Williams2024paradigm}.

%Taken together, our study extends ongoing conversations about human–AI collaboration in qualitative research~\cite{jiang2021serendipity, InfraNodus2025, chew2024aiqualresearch}. It shows that while AI can accelerate some analytic processes, the epistemic commitments of reflexive qualitative inquiry require tools that deliberately scaffold—not replace—reflexivity, transparency, and interpretive depth.

%In this discussion we have shown how DeTAILS advances existing AI-assisted analysis tools by combining efficiency with structured opportunities for researcher engagement. Our results triangulate across qualitative and quantitative data: participants completed coding far more quickly, reported low workload and high manageability, and yet still emphasized the importance of verifying, refining, and questioning AI outputs. These dynamics underscore both the strengths and limitations of current approaches. DeTAILS demonstrates that LLMs can scale qualitative analysis without wholly sacrificing interpretive depth, but also that careful design is necessary to avoid displacing reflexivity with correction work or over-trust in polished outputs.

\subsection{Designing for Reflexive and Trustworthy AI Collaboration}

Triangulating across interviews, logs, and surveys, it is clear that participants’ trust grew not from automation alone but from being able to scrutinize, verify, and correct outputs. Quantitative manageability scores and agreement metrics bolster these accounts, showing that trust was scaffolded through transparency and editability. Yet to move closer to reflexive thematic analysis, systems must go beyond facilitating correction: they need to deliberately create space for reflexivity.

Classic criteria of trustworthiness—credibility, dependability, confirmability, and transferability~\cite{LINCOLN1985}—help explain why this matters. High F1 scores and efficiency metrics provide evidence of dependability, but participants’ concerns about nuance and voice remind us that credibility and confirmability still rest on human reflexivity, not machine alignment. Similarly, AttrakDiff scores showing strong manageability do not ensure that interpretive depth has been preserved. These findings underscore Morrow’s~\cite{Morrow2005} and Finlay’s~\cite{Finlay2002} view that reflexivity is not optional but foundational: researchers must continuously interrogate their own positionality and the power relations shaping analysis.

Our study showed that DeTAILS prompted reflexivity, but primarily about the AI’s suggestions rather than researchers’ own interpretive assumptions. This is not reflexivity in the RTA sense~\cite{BraunANDClarke2019}, but a more pragmatic checking of outputs. To foster genuine reflexive practice, tools must actively slow analysts down and foreground alternative interpretations. Backward propagation, reflexive memoing, structured pauses, positionality statements, and analytic audit trails~\cite{akkermans2021} could all help shift reflexivity from verifying AI outputs toward interrogating the researcher’s own assumptions.

These design imperatives also speak to broader quality debates. Tracy’s~\cite{Tracy2010} “big tent” criteria call for rich rigour, sincerity, and resonance. Efficiency gains and high agreement may contribute to rigour, but they risk overshadowing sincerity and resonance if outputs are too generic or sanitized. Rather than abandoning LLMs for these tensions, we argue for designing explicitly against them: embedding mechanisms that slow users, surface uncertainty, and preserve the interpretive labour—iterative reading, grappling with ambiguity, holding space for participants’ voices --- that cannot be automated without undermining qualitative inquiry.

\subsection{Implications for Design}

Our findings highlight key tensions between efficiency and reflexivity, trust and skepticism, and researcher agency and automation. While DeTAILS succeeded in accelerating thematic analysis and supporting transparency, participants’ experiences also revealed limits: reflexivity often centred on verifying AI outputs rather than interrogating analytic assumptions, forward-only propagation constrained iteration, and polished outputs risked over-trust. Building on these insights, we translate our results into five design considerations for next-generation qualitative analysis systems.

\paragraph{Support Iterative and Bidirectional Workflows}
DeTAILS reduced repetitive coding work but reinforced a codebook orientation by only allowing edits to propagate forward. Reflexive thematic analysis, by contrast, requires recursive movement between codes, themes, and raw data \cite{BraunANDClarke2006, BraunANDClarke2019}. Future systems should support both forward and backward propagation so that refinements at later stages cascade to earlier ones. This would preserve the recursive character of reflexive analysis and scaffold more deliberate opportunities for reflection.

\paragraph{Make Uncertainty Visible}
Quantitative results showed strong agreement scores (F1 up to 1.0) and low workload ratings, but participant interviews revealed skepticism and extensive time spent correcting outputs. This tension highlights the “canny mountain” problem, where AI outputs appear polished and thus invite over-trust, particularly for novices. Designers should resist over-automation and instead surface model uncertainty, flag low-confidence outputs, and provide scaffolds that encourage critical engagement \cite{jiang2021serendipity, hope2025llmqualitativeuses}.

\paragraph{Give Researchers Control Over AI Involvement}
Participants frequently described wanting to disable or bypass the AI when outputs felt intrusive or generic. Mandatory AI assistance risks displacing reflexivity onto model suggestions rather than the researcher’s own analytic lens. Future tools should include opt-in/opt-out controls at each stage, clear provenance markers distinguishing human from AI contributions, and options to advance without invoking the model~\cite{chew2024aiqualresearch, Hitch2024}.

\paragraph{Scaffold Reflexive Practice Deliberately}
Our study showed that participants’ reflexivity often centred on verifying the LLM’s suggestions rather than interrogating their own assumptions. While this still prompted reflective engagement, it diverges from the deeper positional reflexivity emphasized by Braun and Clarke~\cite{BraunANDClarke2019, Finlay2002}. Future systems should embed deliberate reflexive prompts—for example, space to record positionality statements, moments to reflect on analytic choices, or checkpoints asking why particular codes or themes matter. Such scaffolds could help ensure reflexivity remains researcher-driven rather than model-driven.

\paragraph{Adapt to Analytic Styles with Transparency}
We observed wide variation in preferences for code granularity, theme phrasing, and merging strategies. Some participants valued fine-grained codebooks, others preferred higher-level clusters. Current systems, including DeTAILS, treat such preferences as static. Future tools should adapt over time to an analyst’s style, but with safeguards: adaptation must remain transparent, reversible, and always under explicit researcher control. This balance would allow tools to reduce repetitive work while respecting the diversity of qualitative practices~\cite{Nowell2017, saldana2021}.

\section{Limitations and Future Work}
Our study was short-term and conducted in a controlled setting, with participants analysing small Reddit datasets during single sessions. Although many participants expressed interest in applying DeTAILS to their own projects, longer-term deployments are needed to understand how the toolkit integrates into ongoing research and how it shapes analytic rigour over time. Such work would also clarify whether our findings hold for richer qualitative materials such as interviews, field notes, or survey responses~\cite{creswell2016qualitative,denzin2011sage}, which may raise new requirements for preprocessing, navigation, and code review. These forms of data also demand deeper reflexivity than short sessions allow, making longitudinal studies particularly important.

We also did not compare DeTAILS directly with widely used qualitative analysis software such as NVivo or Atlas.ti. These represent the baseline for many research communities, and a head-to-head comparison would clarify where an LLM-supported workflow adds the most value, as well as which features from traditional tools might be incorporated. Relatedly, we observed no statistically significant differences between novice and expert participants. This null result likely reflects the small sample and short study design, and should not be read as evidence that the system equalizes expertise. Prior work cautions that novices are especially vulnerable to over-trust in fluent AI outputs~\cite{sun2024trustllm, gao2024collabcoder}. Future research should therefore investigate how expertise interacts with LLM-supported analysis in sustained use, and how systems can provide scaffolds that support novices’ engagement without encouraging uncritical acceptance.

Finally, our implementation explored only a narrow range of prompting strategies and model types. We relied on a general-purpose LLM with a chain-of-thought workflow, without testing alternatives such as ReAct prompting~\cite{yao2022react}, reasoning-intensive prompting~\cite{sahoo2024systematic}, or domain-specific fine-tuned models~\cite{depaoli2024inductive}. Different configurations may produce markedly different results, not only in accuracy but also in how reflexivity and trust are shaped during analysis. Developing guidelines for pairing analytic tasks with appropriate models, and for aligning prompt strategies with qualitative epistemologies, remains an important direction for future work.

This study was also conducted in mid-2025, during a period of rapid advances in LLM capabilities. Some of the technical limitations we observed may be addressed quickly as models evolve, while new challenges will undoubtedly emerge. Nevertheless, we believe the core findings of this paper—around reflexivity, trust, agency, and the role of human interpretive labour—remain significant. They point to enduring questions about how qualitative research can engage with AI systems, and how design can ensure that these systems scale analysis without eroding its interpretive commitments.

\section{Conclusion}

We presented DeTAILS, a toolkit for human–LLM collaboration in thematic analysis. DeTAILS integrates LLM support across analytic phases while keeping the researcher in control through features such as persistent memory snapshots, redo-with-feedback loops, and fully inspectable codes and themes. These design choices foreground researcher judgment, promote transparency, and enable direct verification of outputs against raw data.

Our study with 18 qualitative researchers showed that DeTAILS substantially accelerated coding and theming without displacing interpretive depth. Participants completed analyses quickly, achieved high agreement with generated themes, and reported feeling “in the driver’s seat” with the AI positioned as a supportive assistant. Trust was not assumed but earned through iterative checkpoints and editable outputs, highlighting how workflow design can foster agency and reflexivity.

This work makes three contributions: a toolkit that demonstrates how LLMs can scaffold large-scale analysis; an empirical study that reveals both benefits and tensions in human–AI collaboration; and design guidelines for future systems that balance acceleration with interpretive authority. More broadly, our findings show that LLMs can extend—but not replace—qualitative inquiry. By coupling computational scale with researcher reflexivity, DeTAILS illustrates how AI tools can amplify interpretive practice while preserving the epistemological commitments that make qualitative research valuable.

\bibliographystyle{ACM-Reference-Format}
\bibliography{sample-base}

\appendix









\end{document}